\documentclass[fleqn,usenatbib,letters]{mnras}

\usepackage{newtxtext,newtxmath}

\usepackage[T1]{fontenc}

\usepackage{graphicx}
\usepackage{natbib}
\usepackage{amsmath,color,xspace,bigints,comment}
\usepackage{pdflscape}

\title[Constraining $M_{H_2}$ at $z\sim8$]{New constraints on the molecular gas content of a $\mathbf{z\sim8}$ galaxy from JVLA CO(J=2-1) observations}
\author[G. C. Jones et al.]{
Gareth C. Jones$^{1}$\thanks{E-mail: gareth.jones@physics.ox.ac.uk},
Joris Witstok$^{2,3}$,
Alice Concas$^4$,
Nicolas Laporte$^{2,3}$
\\
$^1$Department of Physics, University of Oxford, Denys Wilkinson Building, Keble Road, Oxford OX1 3RH, UK\\
$^2$Kavli Institute for Cosmology,University of Cambridge, Madingley Road, Cambridge CB3 0HA, UK\\
$^3$Cavendish Laboratory,University of Cambridge,19 JJ Thomson Avenue, Cambridge CB3 0HE, UK\\
$^4$European Southern Observatory, Karl-Schwarzschild-Strasse 2, D-85748 Garching bei M\"unchen, Germany
}

\date{Received X / Accepted Y}

\pubyear{2023}

\begin{document}
\label{firstpage}
\pagerange{\pageref{firstpage}--\pageref{lastpage}}
\maketitle

\begin{abstract}
As the primary fuel for star formation, molecular gas plays a key role in galaxy evolution. A number of techniques have been used for deriving the mass of molecular reservoirs in the early Universe (e.g., [CII]158\,$\mu$m, [CI], dust continuum), but the standard approach of CO-based estimates has been limited to a small number of galaxies due to the intrinsic faintness of the line. We present Jansky Very Large Array (JVLA) observations of the $z\sim8.31$ galaxy MACS0416\_Y1, targeting CO(2-1) and rest-frame radio continuum emission, which result in upper limits on both quantities. Adding our continuum limit to the published far-infrared (FIR) spectral energy distribution (SED), we find a small non-thermal contribution to the FIR emission, a low dust mass ($\rm\log_{10}(M_D/M_{\odot})\sim5$), and an abnormally high dust temperature ($\rm T_D\gtrsim90\,K$) that may indicate a recent starburst. Assuming a low metallicity ($Z/Z_{\odot}\sim0.25$), we find evidence for $M_{\rm H_2,CO}\lesssim10^{10}$\,M$_{\odot}$, in agreement with previous [CII] investigations ($M_{\rm H_2,[CII]}\sim10^{9.6}$\,M$_{\odot}$). Upcoming JWST observations of this source will result in a precise determination of $Z$, enabling better constraints and an unprecedented view of the gaseous reservoir in this primordial starburst galaxy.
\end{abstract}


\section{Introduction}\label{intro}

Molecular gas reservoirs in galaxies act as potential fuel for future star formation. This gas may be accreted from cosmic filaments or through collisions with gas-rich companions (i.e., `wet' mergers). Recent observations suggest that the cosmic density of molecular gas increased from early times until a peak between $z\sim3$ and $z\sim1$ and decreased to the present day (e.g., \citealt{deca19,arav23,boog23}). This evolution matches the cosmic density of star formation rate (e.g., \citealt{bouw20}), suggesting that star formation in early galaxies ($z>4$, or $\lesssim1.5$\,Gyr after the Big Bang) was powered by gas accreted from past mergers and inflows. 

 Since H$_2$ lacks a permanent dipole moment, molecular gas is difficult to observe directly. Instead, we must rely on indirect tracers. The most commonly used tracer has been carbon monoxide (CO), which is the second most abundant molecule ($N_{CO}/N_{H_2}\sim10^{-4}$) and features a host of rotational emission lines (i.e., $J\rightarrow J-1$; see review of \citealt{bola13}). By observing CO(1-0), the molecular gas mass of a galaxy may be found by assuming a mass-to-light ratio (i.e., $\rm M_{H_2}=\alpha_{CO}L'_{CO(1-0)}$). This ratio $\alpha_{CO}$ has standard values for starburst ($\sim0.8$\,M$_{\odot}$\,$\rm (K\,km\,s^{-1}\,pc^2)^{-1}$) and Milky Way-like systems ($\sim4.36$\,M$_{\odot}$\,$\rm (K\,km\,s^{-1}\,pc^2)^{-1}$), as well as metallicity-dependent forms (e.g., \citealt{nara12}). While other tracers based on line emission (e.g. [CI], [CII], HD; \citealt{walt11,zane18,jone20}), dust continuum emission (e.g., \citealt{deca22,eale23,hash23}), and dust attenuation (e.g. \citealt{guve09,brin13,conc19}) have risen to wide use, CO still acts as the gold standard molecular gas mass estimator.

Due to its faintness with respect to [CII], detection of CO at $z>4$ requires long exposure times and/or exceptionally bright targets. Despite this, there have been a multitude of CO detections between $4<z<6$ in bright starburst galaxies (e.g., \citealt{hodg12,jime18,pave18,riec21a,riec21b,zava21,viei22,eale23,fria23,stan23}), quasar hosts (e.g., \citealt{riec06}), and normal star-forming galaxies (SFGs; e.g., \citealt{dodo18,pave19,garc22,lee23}). Studies of CO within the epoch of reionisation (EoR; $z\gtrsim6$ have been more limited, with CO detections in bright galaxies up to $z\sim 7.5$ (e.g., \citealt{jaru21,ono22, deca22, deca23, feru23}).
In this work, we detail the results of some of the first observations of CO at $z>8$, targeting the $z\sim8.31$ galaxy MACS0416\_Y1.

MACS0416\_Y1 was originally selected as a $z>8$ Lyman break candidate behind the lensing cluster MACS0416 \citep{lapo15}. This source is gravitationally lensed by the cluster, resulting in a magnification of $\mu\sim1.43$ (\citealt{kawa16})\footnote{We will assume this factor throughout this work.}. 
It has been detected in continuum emission using multiple Hubble Space Telescope (HST) Wide Field Camera 3 (WFC3) filters (e.g., F125W, F140W, F160W), the Ks band of the Very Large Telescope (VLT) High Acuity Wide field K-band Imager (HAWK-I), channel 2 of the Spitzer Infrared Array Camera (IRAC), and the Atacama Large Millimetre/submillimetre Array (ALMA) 850$\mu$m (\citealt{lapo15,bram16,tamu19}). 

It features both strong [CII]\,$158\,\mu$m (L$_{[CII]}=(1.40\pm0.22)\times10^8$\,L$_{\odot}$, \citealt{bakx20}) and [OIII]\,$88\,\mu$m (L$_{[OIII]}=(13\pm3)\times10^8$\,L$_{\odot}$; \citealt{tamu19}) emission, implying a star formation rate SFR$\sim60$\,M$_{\odot}$\,year$^{-1}$. By combining a ultraviolet to far-infrared (UV-FIR) spectral energy distribution (SED) model and a stellar evolution model, \citet{tamu19} estimate $M_{H_2}\sim10^{10}$\,M$_{\odot}$ and $M_*\sim10^{9.5}$\,M$_{\odot}$. Using the star-forming main sequence (MS) formulations derived from observations of $0<z<6$ galaxies (e.g., \citealt{spea14, pope23}), an MS galaxy with this stellar mass at this redshift is expected to feature a SFR$\sim25$\,M$_{\odot}$\,year$^{-1}$ (in agreement with the $5<z<10$ results from simulations; e.g., \citealt{dsil23}). Because MACS0416\_Y1 features a higher SFR, it is likely a starburst galaxy. A low-resolution observation of the [CII] velocity field shows evidence for a rotational gradient (\citealt{bakx20}), which is resolved into three star-forming clumps with a beam-deconvolved spatial extent of $0.5''\times0.3''$ in high-resolution [OIII] observations \citep{tamu23} that are also seen in rest-frame UV photometry. 

Using the Jansky Very Large Array (JVLA), we observed this object in CO(2-1) emission and rest-frame radio continuum emission. While this resulted in a non-detection, it enables us to place new constraints on the molecular gas content, non-thermal emission, and dust properties in this high-redshift source. We use a standard concordance cosmology ($h_o$, $\Omega_m$, $\Omega_{\Lambda}$ = 0.7, 0.3, 0.7) throughout, where $1''\sim4.7$\,kpc at $z\sim8.31$.

\section{Data calibration and imaging}\label{datadesc}
Our observations were taken with the JVLA in A-configuration over nine 3\,hr executions between 24 December 2020 - 29 January 2021 under the partially completed project 20B-194. The bandpass and flux calibrator was 3C147, while J0416-1851 was used as a phase calibrator. In order to target CO(2-1) at 24.758\,GHz, we used two basebands in K-band (20.488-21.512\,GHz and 24.488-25.512\,GHz), each with eight spectral windows (SPWs) of 64\,channels of 2\,MHz. 

Due to an error in scan intent declarations, the flux and bandpass calibration scans were unusable, so an extra 30\,minute scan of the phase calibrator was taken. Since J0416-1851 is a high-quality (`P'-grade) calibrator, we were able to use these observations to determine its flux and thus use it as a flux/bandpass calibrator for the previous scans.

The nine executions were calibrated using the Common Astronomy Software Applications package (CASA; \citealt{casa22}) using a standard high-frequency calibration pipeline, but with a manual flux calibration (i.e., no CASA fluxscale) due to the observation setup. The resulting visibilites were inspected, and additional flagging of bad antennas and edge flagging was performed before the pipeline was re-run. One execution was excluded because of significant calibration issues. 

Continuum images were created using CASA tclean with the line-free SPWs of all acceptable executions in multifrequency synthesis (MFS) mode with natural weighting. This results in a synthesised beam of $0.19''\times0.09''$ at $180.0^{\circ}$ and root mean square (RMS) noise level of $1.9\,\mu$Jy\,beam$^{-1}$. 

The basebands of each execution that contained redshifted CO(2-1) emission were separated (CASA split), combined into a single measurement set (CASA concat), and imaged (CASA tclean in `cube' mode, natural weighting, 6\,MHz channels), resulting in a cube with a mean synthesised beam of $0.17''\times0.09''$ at $180.1^{\circ}$ and a mean RMS noise level per channel of $\sim30\,\mu$Jy\,beam$^{-1}$. Since no continuum emission is detected (see Section \ref{contsec}), we do not perform continuum subtraction.

\section{Analysis}\label{analysis}

\subsection{Continuum}\label{contsec}
\begin{figure}
\centering
\includegraphics[width=0.48\textwidth]{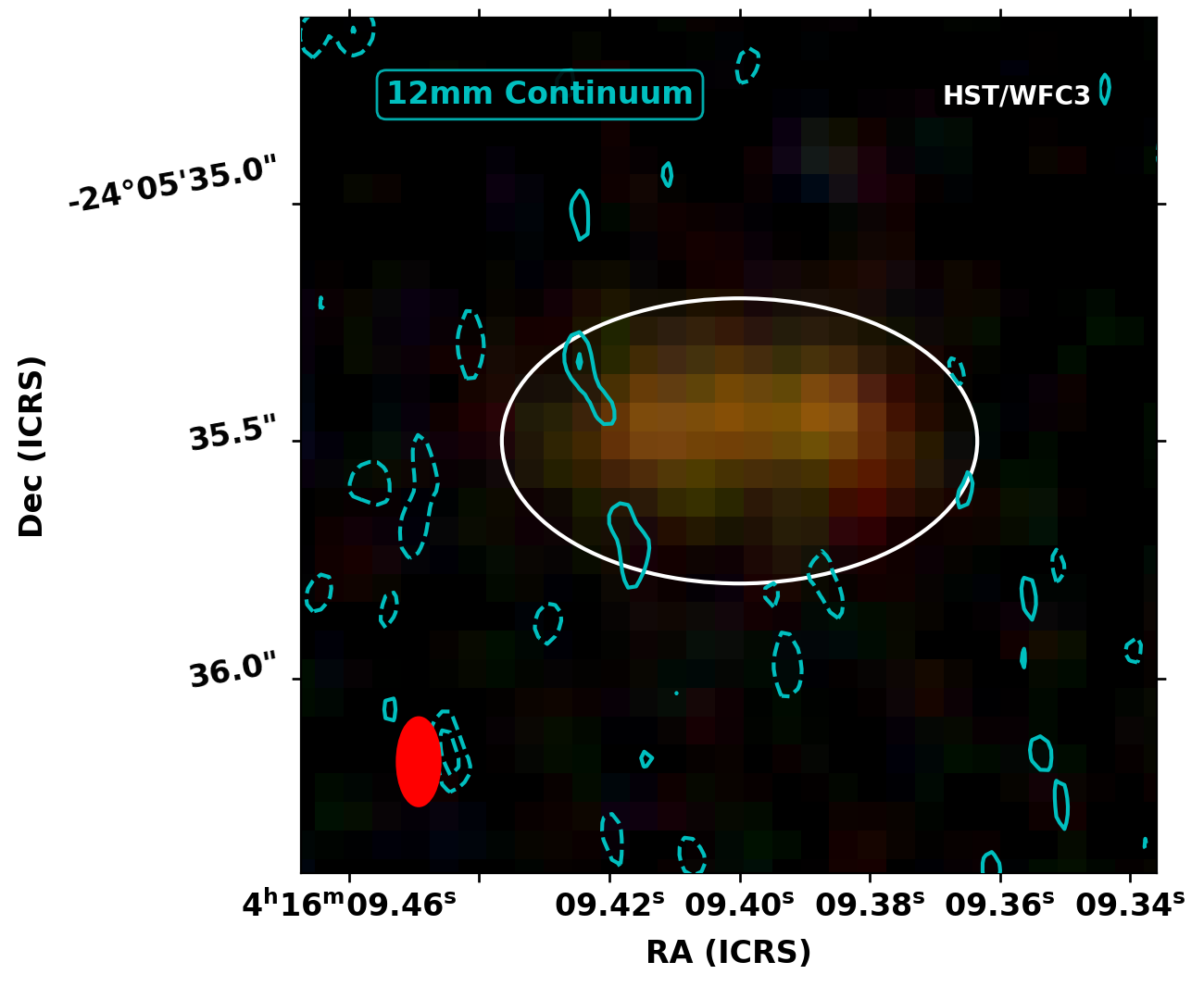}
\caption{Continuum emission ($\nu_{obs}=25$\,GHz) of MACS0416\_Y1 (cyan contours), where contours are shown at $\pm2,3,4,\cdots\times\sigma$ ($1\sigma=1.9\,\mu$Jy\,beam$^{-1}$). For reference, we include a three-colour image of HST/WFC3 (F105W/F125W/F140W). The synthesised beam is represented by a filled red ellipse to the lower left, while the assumed aperture of MACS0416\_Y1 is given as a hollow white ellipse.}
\label{contimage}
\end{figure}

No significant emission is detected in our continuum image ($\nu_{obs}\sim25$\,GHz), as seen in Figure \ref{contimage}. 
To place a conservative estimate on the continuum emission, we use a large elliptical aperture of double the deconvolved [OIII]$88\mu$m full width at half maximum (FWHM) of \citet{tamu19}, resulting in $1.0''\times0.6''$. This implies a $3\sigma$ upper limit of $S_{12mm}<28\,\mu$Jy. Here, we combine this upper limit with archival continuum flux density values to examine the constraints on dust properties and SFR that these implies.

Only three FIR continuum observations have been published to date: $\rm S_{850\,\mu m}=137\pm26\,\mu$\,Jy and $\rm S_{1.5\,m m}<18\,\mu$\,Jy from \citet{tamu19}, and $\rm S_{1.14\,m m}<174\,\mu$\,Jy from \citet{bakx20}, where each limit is $3\sigma$. Several other ALMA programs have targeted the $850\,\mu$m and 1.14\,mm emission, so future works may combine them to place tighter estimates on each flux density. However, only one program has targeted this source at $\sim3.2$\,mm (band 3, 2021.1.00075.S; PI Ono). We apply the ALMA staff calibration (i.e., ScriptForPI.py) to the raw data and create a $3.2\,$mm continuum image using CASA tclean with MFS mode, natural weighting, and a conservative frequency range to exclude possible line emission. This results in a continuum non-detection, and a $3\sigma$ upper limit of $\rm S_{3.2\,mm}<21\,\mu$\,Jy.

We begin by adopting the modified blackbody (MBB) model of \citet{carn19}, which includes the cosmic microwave background (CMB) corrections of \citet{dacu13} and makes no assumption on optical depth. If we assume the dust is emitted from approximately the same area as the [OIII] emission and adopt a dust absorption coefficient of $\kappa_o=0.04$\,m$^2$\,kg$^{-1}$ at a critical density of 250\,GHz (\citealt{beel06}), then this model is reduced to three free parameters: the dust mass ($\rm M_D$), dust temperature ($\rm T_D$), and dust emissivity index ($\beta_{IR}$). To extend this model to lower frequencies, we include the non-thermal emission (i.e., combined synchrotron and free-free) model of \citet{alge21}, which also only has three free variables: the synchtrotron slope ($\alpha_{NT}$), a normalisation factor ($S_{\nu'}$), and the fraction of flux from the free-free component ($f_{th}$). 

\begin{figure*}[h!]
\centering
\includegraphics[width=\textwidth]{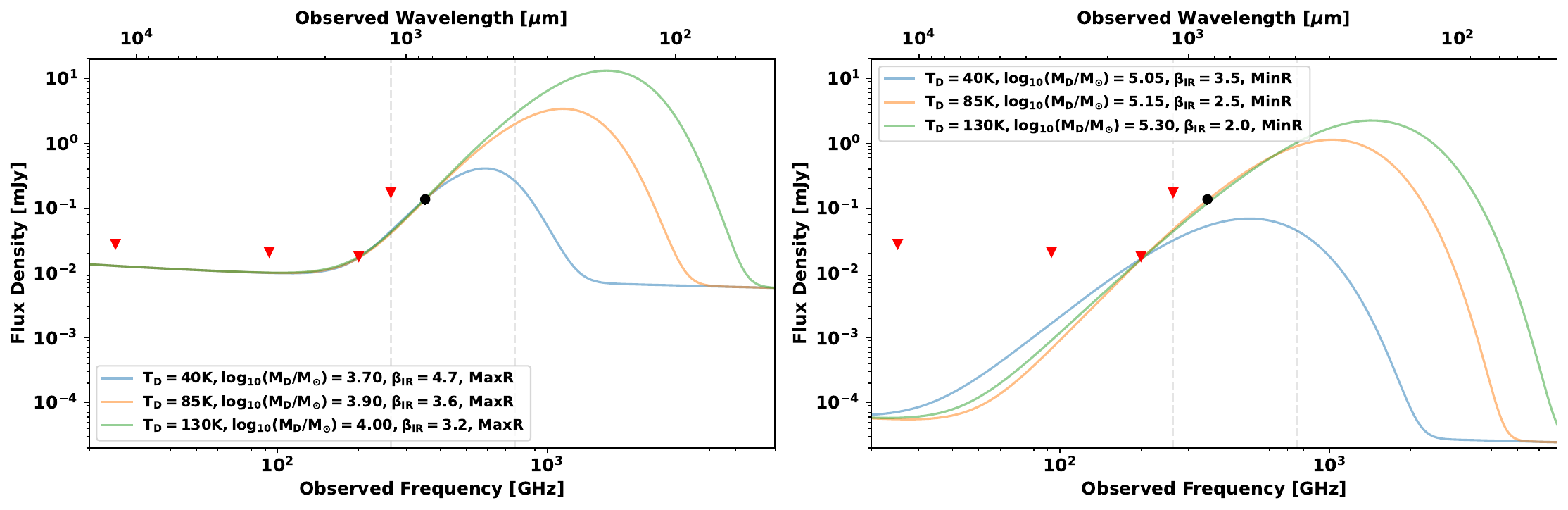}
\caption{Radio-FIR SED of MACS0416\_Y1. We include our upper limit of the 25\,GHz flux density, as well as a detection and limits from \citet{tamu19}, \citet{bakx20}, and archival data (band 3, 2021.1.00075.S; PI Ono). The detection is shown as a black circle with $1\sigma$ error bars, while the upper limits are shown at $3\sigma$ as downward-facing red triangles. The left panel shows the assumption of maximum non-thermal contribution, while the right panel includes the non-thermal contribution assuming SFR$\sim100$\,M$_{\odot}$\,year$^{-1}$. In each panel, we show illustrative models that meet the detections and limits. Dashed vertical lines denote the integration range for L$_{\rm FIR}$ ($\lambda_{rest}=42.5-122.5\,\mu$m).}
\label{sedfig}
\end{figure*}

With only one detection and four upper limits, we cannot place constraints on all six free parameters of this model simultaneously. Instead, we may explore the constraints that our radio and FIR points give. First, we assume standard values for $\alpha_{NT}=0.8$ and $f_{th}=0.1$ (e.g., \citealt{cond92,alge21}), and normalise the non-thermal emission so that our continuum limits are met (see `MaxR' results in left panel of Figure \ref{sedfig}). This results in a limit of $\rm S_{1.4\,GHz}<120\,\mu$\,Jy, or a SFR$_{\rm 1.4GHz}$ limit of $\lesssim2.5\times10^4$\,M$_{\odot}$\,year$^{-1}$ (\citealt{cond92}). This is much greater than the expected SFR of this object (SFR$\sim60-100$\,M$_{\odot}$\,year$^{-1}$; \citealt{bakx20}), so this limit is not highly constraining. Since the radio continuum point is not informative, we may assume a smaller SFR$\sim10^2$\,M$_{\odot}$\,year$^{-1}$, which implies a much smaller radio contribution (see `MinR' results in right panel of Fig. \ref{sedfig}).

In both cases of radio emission (`MaxR' and `MinR'), we examine the FIR portion of the SED by assuming a dust temperature (40\,K, 85\,K, or 130\,K) and explore what $\rm M_D$ and $\beta_{IR}$ values are required for a given dust temperature to satisfy the $S_{850\,\mu m}$ detection and $S_{1.5\,m m}$ non-detection. We find that a larger non-thermal contribution requires a smaller dust mass and steeper spectrum (i.e., higher $\beta_{IR}$). On the other hand, a higher dust temperature requires a larger dust mass and shallower slope, and results in a higher FIR luminosity (area between the dashed vertical lines) and thus SFR$_{FIR}$.

Since this SED only contains a single point, these models are for illustration only. But given the SFR of this source and the fact that $\beta_{IR}$ values greater than $\sim2.5$ are rarely seen (e.g., \citealt{wits23}), it is likely that: i.) Non-thermal emission does not significantly contribute to the FIR luminosity, ii.) The dust temperature is high ($\gtrsim 90$\,K), and iii.) The dust mass is quite small ($\rm M_D\sim10^5$\,M$_{\odot}$). These last two conclusions are in agreement with the SED analysis of \citet{bakx20}.

Note that these dust masses are smaller than that of \citet[$\rm M_D=4\times10^6$\,M$_{\odot}$]{tamu19}, who assumed $\beta_{IR}=1.5$, $\rm T_D=50$\,K, and a different dust absorption coefficient. The primary difference is their use of a UV-to-FIR SED model that includes dust attenuation and scaled FIR templates, rather than our use of a single MBB with a flexible $\beta_{IR}$ and $\rm T_D=50$\,K. We are unable to recreate a model that uses $\beta_{IR}=1.5$ and $\rm T_D=50$\,K and obeys $\rm S_{850\,\mu m}=137\pm26\,\mu$\,Jy and $\rm S_{1.5\,m m}<18\,\mu$\,Jy, suggesting that a future flexible UV-to-FIR SED model is required.

Briefly, we note that the derived luminosity-weighted dust temperature limit implied by our exploration ($\rm T_D\gtrsim 90$\,K) is much higher than commonly used mass-weighted dust temperatures at low-redshift ($\rm T_D=25$\,K; \citealt{scov16}), as well as other high-redshift luminosity-weighted values (e.g.; $\rm T_D\sim 40-70$\,K at $z\sim5-7$, \citealt{bakx21,ikar22,jaru21,mann22,viei22}). While dust temperature has been found to increase with redshift (e.g., \citealt{bouw20,wits23,jone23}), most correlations would predict $\rm T_D\sim 60-70$\,K at $z=8.31$. The fact that we predict a higher temperature could be interpreted in multiple ways: i.) MACS0416\_Y1 may contain abnormally warm dust due to its high specific SFR (e.g., \citealt{lian19,mits23}), ii.) The correlation between redshift and dust temperature is exponential rather than linear (e.g., \citealt{vier22}), or iii.) The current dataset (a single point and multiple upper limits) and/or model (modified blackbody) are insufficient to describe the dust properties. Since many studies of dust properties at high-redshift use a single FIR continuum detection and assume a dust temperature or template (e.g., \citealt{beth20,fuda21,bowl23}), more observations are required to resolve this ambiguity.

\subsection{Molecular gas}\label{mh2sec}
\begin{figure*}
\centering
  \raisebox{-0.5\height}{\includegraphics[width=0.54\textwidth]{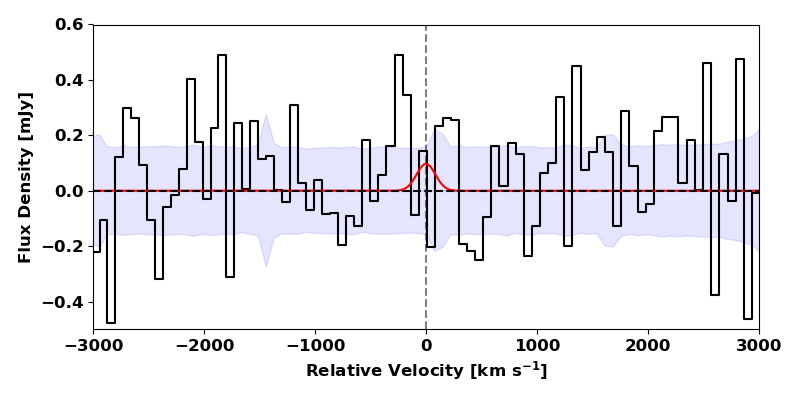}}
  \hspace*{.0in}
  \raisebox{-0.5\height}{\includegraphics[width=0.45\textwidth]{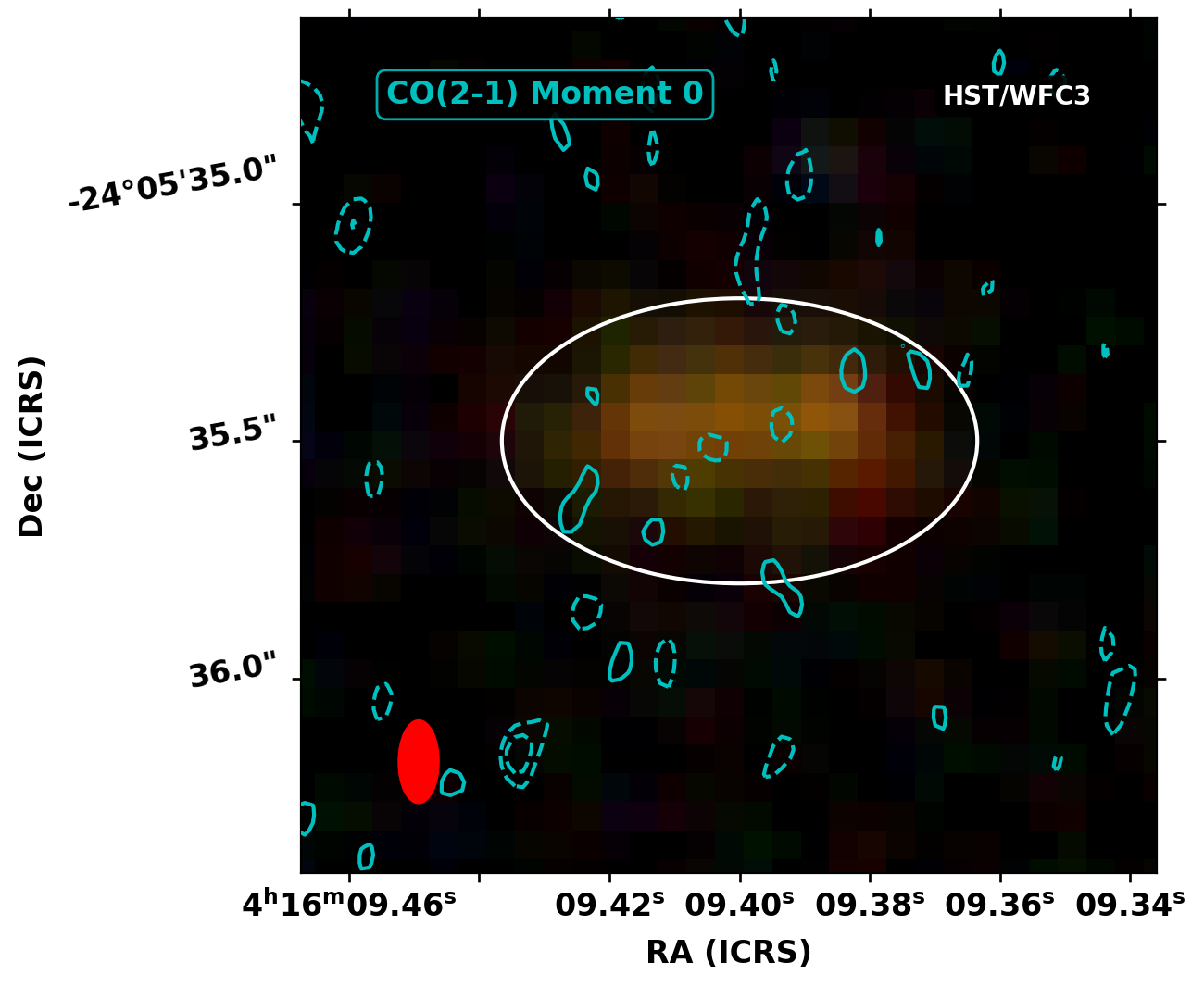}}
\caption{Observed CO(2-1) properties from our JVLA observations. The left panel shows a spectrum extracted from a $1.0''\times0.6''$ ellipse centred on the position of MACS0416\_Y1 (black lines) with associated $1\sigma$ uncertainty (blue shaded region). We also show the expected CO(2-1) line profile for a $10^{10}$\,M$_{\odot}$ gaseous reservoir (red profile; assuming FWHM$=200$\,km\,s$^{-1}$ and $\alpha_{CO}=0.8$), which is neither confirmed nor excluded by our data. The right panel is a CO(2-1) moment 0 map for $v=[-100,100]$\,km\,s$^{-1}$ of MACS0416\_Y1 (cyan contours), where contours are shown at $\pm2,3,4,\cdots\times\sigma$ ($1\sigma=4.5\,$mJy\,beam$^{-1}$). For reference, we include a three-colour image of HST/WFC3 (F105W/F125W/F140W). The synthesised beam is represented by a filled red ellipse to the lower left, while the assumed aperture of MACS0416\_Y1 is given as a hollow white ellipse.}
\label{M0SPEC}
\end{figure*}

A spectrum extracted from our combined data cube using an ellipse centred on the expected location of MACS0416\_Y1 is shown in Figure \ref{M0SPEC}. It is clear that the observed spectrum (black line) shows no significant line emission. If there was indeed $10^{10}$\,M$_{\odot}$ of molecular gas, then we would expect the line profile shown by the red Gaussian (peak amplitude=$0.097$\,mJy, FWHM$=200$\,km\,s$^{-1}$; assuming $\alpha_{CO}=0.8$\,M$_{\odot}$\,$(K\,km\,s^{-1}\,pc^2)^{-1}$), which peaks at $<1\sigma$. That is, we lack the sensitivity to confirm or rule out this amount of molecular gas.

To place an upper limit on the integrated flux density, we first collapse the data cube around the expected CO(2-1) frequency, assuming a width of $\sim200$\,km\,s$^{-1}$ (based on FWHM$_{\rm [OIII]88\mu m}$; \citealt{tamu19}). This map (shown in the right panel of Fig. \ref{M0SPEC}) has an RMS noise level of 4.5\,mJy\,beam$^{-1}$\,km\,s$^{-1}$, corresponding to a $3\sigma$ upper limit on the integrated flux density of $S\Delta $v$_{\rm CO(2-1)}<72$\,mJy\,km\,s$^{-1}$. 

Using the standard equation of \citet{solo92}, this results in an upper limit of:
\begin{equation}
\rm M_{H_2}<(2.5\times10^{10}\,M_{\odot})\frac{\alpha_{CO}}{0.8\,M_{\odot}/(K\,km\,s^{-1}\,pc^2)}
\frac{1.43}{\mu}\frac{1}{r_{21}}\frac{0.76}{f_{CMB}}
\end{equation}
where we have assumed a starburst-like $\alpha_{CO}$ and $r_{21}$, as motivated by the high [OIII]\,88$\mu$m/[CII]\,158\,$\mu$m ratio detected by \cite{tamu19} and \cite{bakx20}, as well as the discovery of a likely starburst-driven dust bubble in high-resolution [OIII]\,88$\mu$m imaging \citep{tamu23}. The effect of observing CO against the CMB is taken into account through the factor $\rm f_{CMB}$ (\citealt{dacu13}):
\begin{equation}
\rm f_{CMB}\equiv\frac{S_{CO,observed}}{S_{CO,intrinsic}}=1-\frac{B_{\nu}(T_{CMB})}{B_{\nu}(T_{ex,CO})}
\end{equation}
where $\rm S_{CO}$ is the CO(2-1) flux density and $B_{\nu}(T)$ is a blackbody function. We assume that the dust and molecular gas are in thermal equilibrium ($\rm T_{ex,CO}=T_D\sim90$\,K). 

This dust temperature is the lower limit implied by the SED exploration of Section \ref{contsec}, and a higher $\rm T_D$ results in a higher $f_{CMB}$ and thus smaller limit on $\rm M_{H_2}$ (e.g., $f_{CMB}=0.84$ for $\rm T_D=130$\,K, yielding a limit of $\rm M_{H_2}<2.2\times10^{10}$\,M$_{\odot}$). These temperatures are quite high, but a derivation of $T_{ex,CO}$ requires multiple CO detections and excitation modelling (e.g., \citealt{dadd15}). If we assume $\rm T_D=90$\,K, but Milky Way-like values of $\alpha_{CO}\sim4.3$ and $\rm r_{21}\sim0.5$ (e.g., \citealt{cariw13}), this results in a much more conservative limit of $\rm M_{H_2}<2.6\times10^{11}$\,M$_{\odot}$.

Previous [CII] observations yielded a luminosity of L$_{[CII]}=(1.40\pm0.22)\times10^8$\,L$_{\odot}$ (\citealt{bakx20}). This line has proven to be a reliable tracer of SFR in galaxies at low and high redshift (e.g., \citealt{delo14,scha20}), although some studies have found that it acts as a reliable tracer of molecular gas (e.g., \citealt{zane18,madd20,gurm23}). While the $L_{[CII]}$-M$_{H_2}$ relation may represent a linked relation between $L_{[CII]}$-SFR and SFR-M$_{H_2}$ (i.e., the Kennicutt-Schmidt relation; \citealt{kenn98,vizg22}), it is clear that [CII] traces molecular gas that is faint in CO emission (i.e., `CO-dark' gas; \citealt{hu21}). Using $\alpha_{[CII]}\sim30$\,M$_{\odot}$\,L$_{\odot}$$^{-1}$ (\citealt{zane18}) results in an estimated $M_{H_2}\sim4.2\times10^9$\,M$_{\odot}$ for MACS0416\_Y1. 

We may also use the predicted dust mass of Section \ref{mh2sec} ($\sim10^5$\,M$_{\odot}$) to estimate M$_{H_2}$ by adopting the dust-to-gas ratio ($\delta_{DGR}$) prescription of \citet{li19}:
\begin{equation}
\log_{10}\left(\delta_{DGR}\right) = (2.445\pm0.006)\log_{10}\left(Z'\right)-(2.029\pm0.003)
\end{equation}
where $Z'=Z/Z_{\odot}$. If we assume $Z'\sim0.25$ (\citealt{bakx20}), this results in $\delta_{DGR}\sim10^{-3.5}$, and a predicted gas mass of $M_{H_2}\sim10^{8.5}$\,M$_{\odot}$. Since this relation was derived using a sample of star-forming main sequence galaxies at $z\sim0-6$ and MACS0416\_Y1 is likely a starbursting galaxy at higher redshift, it is conceivable that MACS0416\_Y1 may feature an even smaller $\delta_{DGR}$ value (as seen in other high-redshift sources; e.g., \citealt{hash23}), and consequently a higher $M_{H_2}$.

A [CII] velocity gradient was detected in this source, which was assumed to be rotation, resulting in a dynamical mass estimate of $M_{dyn}=(1.2\pm0.4)\times10^{10}$\,M$_{\odot}$ (\citealt{bakx20}). Since high-resolution [OIII] observations revealed that this source is composed of several clumps with complex kinematics, this value may only be used as an approximate estimate. However, since the stellar mass of this object is approximately $M_*\sim10^9$\,M$_{\odot}$ (\citealt{tamu19}) and we find a small dust mass ($\rm M_D\sim10^5$\,M$_{\odot}$), there may be room in the mass budget for a large amount of gas ($\sim10^{10}$\,M$_{\odot}$).

To summarise, the [CII] luminosity of this source implies $M_{H_2}\sim4.2\times10^9$\,M$_{\odot}$, which is in agreement with the mass limits imposed by the non-detection of CO(2-1) and from a dynamical mass decomposition, as well as an approximate dust-based estimate. Cosmological zoom-in simulations of galaxies at $z\sim6-7$ suggest smaller gas masses for a comparable galaxy ($M_{H_2}\sim10^{8.5}$\,M$_{\odot}$, \citealt{vall12,pall17}), while observations of more massive galaxies at $z\sim6-7$ return larger molecular gas masses (e.g., \citealt{ono22,feru23}). It can be clearly seen from the values listed in Table \ref{mh2table} that the available data suggest a broad mass range of $M_{H_2}\sim10^{8.5-11.3}$\,M$_{\odot}$. In order to place a tighter constraint, the gas-phase metallicity is required to calibrate $\delta_{DGR}$ and $\alpha_{CO}$. Luckily, upcoming JWST observations of this source with the integral field unit (IFU) of the Near-InfraRed Spectrograph (NIRSpec; PID 1208, PI Willott) will target MACS0416\_Y1 in both low- ($R\sim100$; Prism) and high-spectral resolution ($R\sim2700$; G395H), which will result in a precise estimate of metallicity through well-tested rest-frame optical line ratios. 

\begin{table}
\centering
\begin{tabular}{c|c}
Method & M$_{H_2}$ [M$_{\odot}$] \\ \hline
L'$_{\rm CO(2-1),SB}$ & $<2.5\times10^{10}$ \\
L'$_{\rm CO(2-1),MS}$ & $<2.6\times10^{11}$ \\
$\rm [CII]$ Luminosity & $\sim4.2\times10^9$\\
$\delta_{DGR}$ & $\sim3.1\times10^8$\\
$\rm [CII]$ Dynamical Decomposition & $<1.1\times10^{10}$
\end{tabular}
\caption{Estimates of M$_{H_2}$ from different tracers.}
\label{mh2table}
\end{table}

\section{Conclusion}\label{conc}
Here, we present an updated analysis of the dust and molecular gas properties in the $z\sim8.31$ galaxy MACS0416\_Y1, including both archival data and new JVLA observations targeting CO(2-1) and rest-frame radio continuum emission. 

Since the continuum emission is not detected at $\nu_{obs}\sim25$\,GHz, we examine the implications this has on the non-thermal emission. Assuming a standard non-thermal fraction and synchrotron slope, our non-detection implies a 1.4\,GHz upper limit of $S_{1.4\,GHz}<120\,\mu$\,Jy. If the \citet{kenn98} scaling law is applied, this results in a SFR$_{\rm radio}\lesssim2.5\times10^4$\,M$_{\odot}$\,year$^{-1}$. This is much larger than the observed SFR, implying an uninformative constraint. Assuming a lower SFR$\sim10^2$\,M$_{\odot}$\,year$^{-1}$ results in $S_{1.4\,GHz}\sim0.5\,\mu$\,Jy, and thus a minuscule contribution of nonthermal emission to the luminosity.

The archival FIR SED (which consists of multiple non-detections and one detection) is re-examined and modelled with a modified blackbody. We find that these data suggest a low dust mass ($\sim10^5$\,M$_{\odot}$) and high dust temperature ($\rm T_D\gtrsim90$\,K), in agreement with past results (\citealt{bakx20}).

Our non-detection of CO(2-1) is used to place a constraint on M$_{H_2}$, which is compared to estimates from [CII] luminosity, dust mass, and [CII] kinematics. While the L$\rm _{[CII]}$-based estimate of the H$_2$ mass ($\sim4.2\times10^9$\,M$_{\odot}$) is in agreement with the CO-based limit ($<2.5\times10^{10}$\,M$_{\odot}$) and a mass decomposition based on M$_{dyn,[CII]}$ ($<1.1\times10^{10}$\,M$_{\odot}$), the dust-to-gas ratio-based estimate is much lower ($\sim3.1\times10^8$\,M$_{\odot}$). If the [CII]-based estimate is accurate, this suggests a higher gas-to-dust ratio than previously expected .

While the constraints in this work are not yet precise, they will be refined greatly in the near future through synergy with high-resolution ALMA [CII], [CI], and CO observations as well as JWST/NIRSpec IFU observations (Witstok et al. in prep). These will result in new estimates of the gas-phase metallicity, non-thermal and FIR continuum emission, SFR, and gas properties in MACS0416\_Y1, enabling a detailed view of this primordial galaxy.

\section*{Data Availability}
The uncalibrated datasets analysed in this work are available from the NRAO data archive (\url{https://data.nrao.edu/portal/#/myDataViewer}) and ALMA data archive (\url{https://almascience.nrao.edu/asax/}). Calibrated data are available from the corresponding author upon reasonable request.

\section*{Acknowledgements}
GCJ acknowledges funding from the ``FirstGalaxies'' Advanced Grant from the European Research Council (ERC) under the European Union’s Horizon 2020 research and innovation programme (Grant agreement No. 789056). 
JW gratefully acknowledges support from the Fondation MERAC, the Science and Technology Facilities Council (STFC), the European Research Council (ERC) through Advanced Grant 695671 (``QUENCH''), and the UK Research and Innovation (UKRI) Frontier Research grant RISEandFALL.
NL acknowledges support from the Kavli foundation.
This paper utilizes data obtained with the ALMA Observatory, under program 2021.1.00075.S. ALMA is a partnership of ESO (representing its member states), NSF (USA) and NINS (Japan), together with NRC (Canada), MOST and ASIAA (Taiwan), and KASI (Republic of Korea), in cooperation with the Republic of Chile. We thank the anonymous reviewer for constructive feedback that strengthened this work.

\bibliographystyle{mnras}
\bibliography{references}

\label{lastpage}
\end{document}